\documentclass[compactaffiliation]{Interspeech}

\interspeechcameraready

\title{Learning More with Less: Self-Supervised Approaches for\\Low-Resource Speech Emotion Recognition}

\author[affiliation={1}]{Ziwei}{Gong}
\author[affiliation={1}, equalcontribution]{Pengyuan}{Shi}
\author[affiliation={1}, equalcontribution]{Kaan}{Donbekci}
\author[affiliation={1}]{Lin}{Ai}
\author[affiliation={1}]{Run}{Chen}
\author[affiliation={2}]{David}{Sasu}
\author[affiliation={1}]{Zehui}{Wu}
\author[affiliation={1}]{Julia}{Hirschberg}

\affiliation{}{Columbia University}{USA}
\affiliation{}{IT University of Copenhagen}{Denmark}

\email{\{zg2272,ps3391,kd2939\}@columbia.edu,\{lin.ai,runchen\}@cs.columbia.edu, dasa@itu.dk, zw2804@columbia.edu, julia@cs.columbia.edu}

\keywords{Speech emotion recognition, Transfer learning, Low-resources languages, Multilinguality}

\usepackage{amsmath}
\usepackage{amssymb}

\begin{document}

\maketitle
\renewcommand{\thefootnote}{\fnsymbol{footnote}}  
\footnotetext{Code released: \url{github.com/lynneeai/Speech_Emotion}} 
\begin{abstract}
    
    Speech Emotion Recognition (SER) has seen significant progress with deep learning, yet remains challenging for Low-Resource Languages (LRLs) due to the scarcity of annotated data. In this work, we explore unsupervised learning to improve SER in low-resource settings. Specifically, we investigate contrastive learning (CL) and Bootstrap Your Own Latent (BYOL) as self-supervised approaches to enhance cross-lingual generalization. Our methods achieve notable F1 score improvements of 10.6\% in Urdu, 15.2\% in German, and 13.9\% in Bangla, demonstrating their effectiveness in LRLs. Additionally, we analyze model behavior to provide insights on key factors influencing performance across languages, and also highlighting challenges in low-resource SER. This work provides a foundation for developing more inclusive, explainable, and robust emotion recognition systems for underrepresented languages.

\end{abstract}

\section{Introduction}
Speech Emotion Recognition (SER) is a fundamental task in speech processing with applications in human-computer interaction, affective computing, and mental health monitoring. The goal of SER is to automatically infer emotional states from speech signals, enabling more natural and adaptive interactions between humans and machines. Early approaches to SER relied on statistical learning methods, including Support Vector Machines (SVMs), k-Nearest Neighbors (k-NNs), Gaussian Mixture Models (GMMs), and Hidden Markov Models (HMMs) \cite{sheikhan2013modular,  alam2013amplitude, yang2017enhanced, chaudhari2016selection, schuller2003hidden}. 
With advances in deep learning, more recent SER systems leverage architectures such as Deep Neural Networks (DNNs), Long Short-Term Memory (LSTMs), Recurrent Neural Networks (RNNs), and Convolutional Neural Networks (CNNs) \cite{yildirim2021modified, liu2021speech, langari2020efficient, jain2020speech}, achieving significant performance improvements by leveraging techniques from Automatic Speech Recognition (ASR) and Speech Understanding (SU) \cite{lin2024improving,wu-etal-2024-multimodal, ai2020new,wu2024silentlettersamplifyingllms}.

Despite these advancements, SER models tend to perform well only in High-Resource Languages (HRLs) where large-scale annotated speech emotion datasets are available. Many Low-Resource Languages (LRLs) lack sufficient labeled data, leading to significant performance disparities across different linguistic and cultural contexts. To address this issue, prior work has explored techniques such as domain adaptation, data augmentation, and transfer learning \cite{hozjan2003context, hassan2013acoustic, song2016cross, zhao2023deep, lin2024improving, wu2025multimodalemotionrecognitionconversations}. While domain adaptation methods attempt to align feature distributions across languages, they often fail to guarantee consistent predictive performance. Data augmentation strategies, such as Generative Adversarial Networks (GANs), require large amounts of high-quality data for stable training, which is often unavailable for LRLs.  This heavy dependence on labeled data limits the scalability of existing approaches and makes them less effective for low-resource scenarios.

Unsupervised learning has been proposed as an alternative to reduce reliance on labeled data, yet it remains a challenging avenue for SER. Contrastive Learning (CL) \cite{li2021contrastive, alaparthi2022scser} and self-supervised methods such as Bootstrap Your Own Latent (BYOL) \cite{grill2020bootstrap, niizumi2021byol} have shown promise in representation learning for speech tasks without labeled supervision \cite{baevski2020wav2vec}. However, their effectiveness in SER is still under-explored, and challenges remain in ensuring robust feature extraction and maintaining meaningful cross-lingual generalization. Unlike tasks such as Automatic Speech Recognition (ASR), where phonetic structures provide relatively stable patterns, emotion expression is highly variable across speakers and cultures, making self-supervised learning particularly difficult in multilingual SER.



We explore the use of unsupervised learning approaches to improve speech emotion recognition in low-resource settings, providing deeper insights into cross-lingual emotion recognition. We proposal two unsupervised learning approaches, Contrastive Learning and Bootstrap Your Own Latent, and demonstrate their potential to enhance cross-lingual generalization. 

Our findings show that these methods significantly improve performance in low-resource settings, achieving F1 score gains of 10.6\% in Urdu, 15.2\% in German, and 13.9\% in Bangla. Also, we analyze model behavior through interpretability methods, including feature visualization, confusion matrix, and post-hoc explainability, identifying key factors that influence performance across languages. Beyond performance improvements, our work highlights persistent challenges in low-resource SER, such as gender bias, linguistic transfer limitations, and dataset imbalances, providing a foundation for future research in fair and effective multilingual speech emotion recognition.
\begin{figure*}
    \centering
    \includegraphics[width=0.95\linewidth]{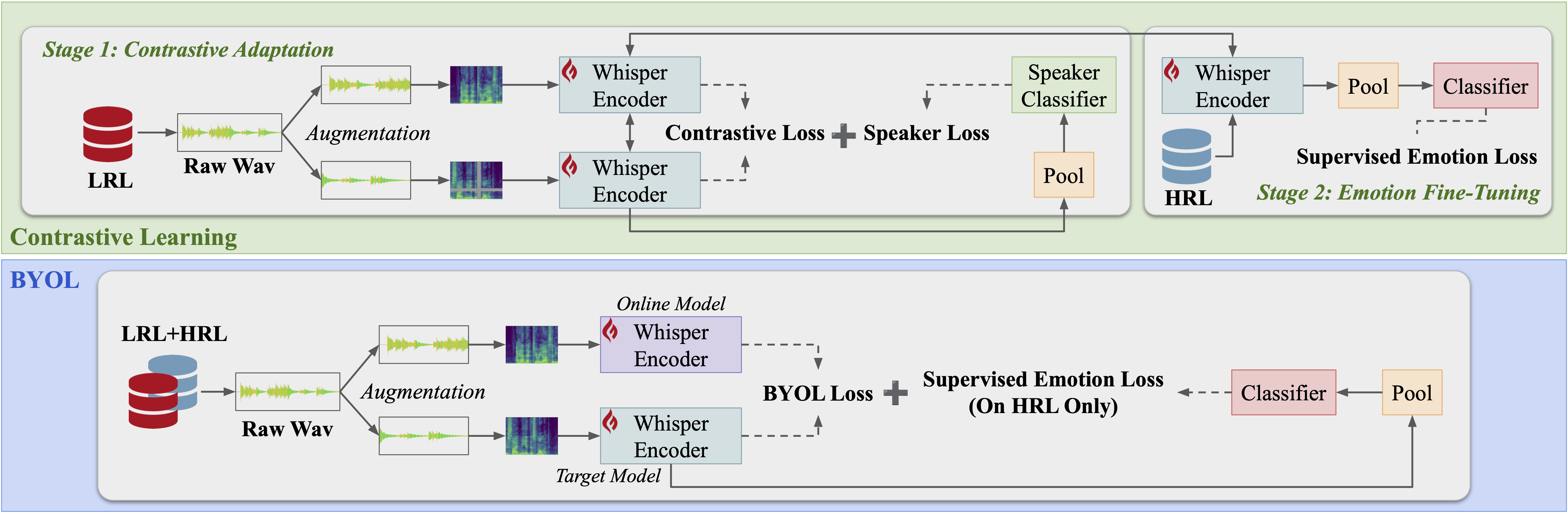}
    \caption{Diagram of our proposed frameworks. Top: Contrastive Learning (CL). Bottom: Bootstrap Your Own Latent (BYOL). HRL and LRL signify High and Low Resource Languages.}
    \label{fig:frameworks}
\end{figure*}

\section{Methodology}
We choose Contrastive Learning (CL) and Bootstrap Your Own Latent (BYOL) for their effectiveness in learning robust, generalizable representations from unlabeled data, a key requirement for low-resource SER.

\textbf{CL} differentiates similar and dissimilar inputs using self-supervised similarity measures, relying on carefully selected negative samples to contrast with augmented positive pairs. CL is more intuitive as it explicitly separates representations but relies on high-quality negative sampling.

\textbf{BYOL}, in contrast, learns representations without negative samples. It maintains two networks—a target and an online network—to ensure training stability and encourage consistency and invariance in learned features.


\subsection{Data Processing and Augmentation} 

For CL, during contrastive speaker adaptation, each utterance has two views: a clean version and an augmented version, forming a positive pair since they share the same speaker label. For fine-tuning and baseline training, only the augmented views are used. We apply several data augmentations, including Gaussian noise injection (SNR sampled uniformly in 10–20 dB range), polarity inversion, gain manipulation, speed perturbation, spectral stretch, and SpecAugment \cite{Park2019-sk}. 

\begin{figure}[t]
    \centering
    \includegraphics[width=0.9\linewidth]{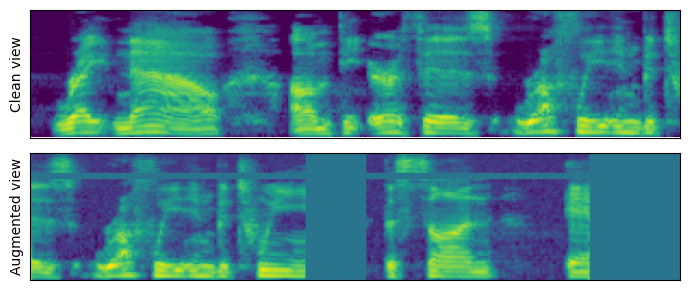}
    \caption{Both views of the same utterance with minimal overlap, forming a positive pair in contrastive adaptation or BYOL.}
    \label{fig:data_preprocess}
\end{figure}

For BYOL, we generate image pairs for training using time-frequency masking, mixUp, and random-resize-crop on mel-spectrograms. These augmentations, applied independently to each view, promote diverse transformations and prevent trivial memorization, fostering robust shared representations. For supervised training (\ref{sec:byol_2}), only time-frequency masking and random-resize-crop are applied with reduced strength, generating three views per sample. 
Figure \ref{fig:data_preprocess} shows an example of resulting spectrograms after augmentation. All feature extraction follows Whisper’s default parameters.
\subsection{Approach 1: Contrastive Learning (CL)}
\label{subsec:contrastive}
The proposed CL approach consists of two stages: speaker-contrastive adaptation and emotion fine-tuning (Figure \ref{fig:frameworks}).
\\
\textbf{Stage 1: Speaker-Contrastive Adaptation.}
We leverage speaker identity, as prior work shows a strong link between SER and speaker recognition (SR). Studies \cite{Ulgen2024-yg, Phukan2023-ns} highlight correlations between speaker embeddings and emotional content, with pre-trained SR models like ECAPA-TDNN \cite{Desplanques2020-jg} outperforming other speech pre-training tasks. However, this connection remains unexplored in cross-lingual transfer learning. We hypothesize that enforcing embedding consistency for LRL speakers prevents the model from learning representations tied only to the HRL, whether lexical or paralinguistic.
We propose a self-supervised contrastive objective on LRL datasets, forming positive/negative pairs based on speaker labels. Since speaker labels are easier to obtain than emotion labels and are commonly included in public datasets, this approach is widely applicable to other LRLs beyond our study.
For the contrastive objective, we use Normalized Temperature-scaled Cross Entropy Loss \cite{Sohn2016-qe}:
\begin{equation}
    l_{i,j} = - \log \frac{\exp(\text{sim}(\mathbf{z}_i, \mathbf{z}_j)/\tau)}{\sum_k \exp (\text{sim}(\mathbf{z}_i, \mathbf{z}_k)/\tau)}
\end{equation}
where $\tau$ is the temperature setting, $(i,j)$ are indices of a positive pair (same speaker label), $\mathbf{z}$ are model embeddings, and $k$ denotes indices with a different speaker label than $i$. We use cosine similarity as the similarity measure $\text{sim}( \cdot ,\cdot)$, computing the loss across all positive pairs $(i,j)$ and averaging it.
To ensure valid positive/negative pairs, each batch includes 16 randomly sampled speakers from LRL datasets, with 4 utterances per speaker (without replacement). This prevents any speaker from dominating and promotes diverse negative pairs.

\textbf{Stage 2: Emotion Fine-tuning.}
In the second stage, only English emotion-labeled HRL datasets are used. The pre-trained embedding model is fine-tuned with a cross-entropy objective for multi-class emotion classification. To address class imbalance, each batch includes an equal number of utterances per emotion class, preventing any single class from dominating.


\subsection{Approach 2: Bootstrap Your Own Latent (BYOL)}
\label{sec:byol_2}
BYOL \cite{grill2020bootstrap} is a self-supervised representation learning method without negative pairs. Instead, it maintains two network branches, an \textit{online} network and a \textit{target} network, which are updated in a momentum fashion from \textit{online} network. We integrate a supervised objective with a self-supervised BYOL-based objective into a single training stage, as illustrated in Figure \ref{fig:frameworks}. Namely, we optimize the following objectives:
\begin{itemize}
    \item \textit{Supervised objective (HRL data):} A cross-entropy (CE) loss is computed on HRL labeled emotion data, leveraging ground-truth emotion classes. 
    \item \textit{Self-supervised objective (HRL+LRL data):} Simultaneously, the BYOL objective -- following the loss formulation and momentum-based weight update as described in \cite{grill2020bootstrap} -- is applied to all available utterances (both HRL and LRL), with labels ignored for HRL data in this component.
\end{itemize}
Hence, the overall objective for each mini-batch is the sum of the CE loss and BYOL loss weighted by a scalar factor $\lambda$:
{\small
\begin{equation}
    \mathcal{L}_{\text{mixed}} = (1 - \lambda) \mathcal{L}_{\text{CE}}(\theta) + \lambda \, \mathcal{L}_{\text{BYOL}}(\theta, \xi),
\end{equation}
}
where $\theta$ and $\xi$ denote the parameters of the online and target networks, respectively.
To address the issue of imbalanced data between HRL and LRL, we sample randomly from each data source at each training step.
During inference, we use the trained target network with the classification head from the HRL supervised training component and evaluate it on LRL datasets.



\section{Experiments}

\subsection{Datasets}
We investigate the effectiveness of the above methods for low-resource SER on 3 LRLs, Urdu, German, and Bangla, and one HRL English as comparison.

For the HRL, We use MSP-Podcast \cite{LotfianUnknown-ah} and IEMOCAP \cite{Busso2008-sk}, both widely used for English SER. MSP-Podcast consists of speech from online podcast recordings, while IEMOCAP contains both acted and improvised dialogues from 10 actors.

For LRLs, we use EMO-DB (German) \cite{Burkhardt2005-iu}, URDU (Urdu) \cite{Latif2018-rb}, and SUBESCO (Bangla) \cite{Sultana2021-vk}. To ensure consistency across datasets, we keep only utterances from 0.5s to 12s with common labels (“happy,” “sad,” “neutral,” “angry”). We merge the “excited” label with “happy,” following \cite{gong-etal-2024-mapping}. 

For our CL approach, we train on the German, Urdu, and Bangla subsets of Common Voice \cite{Ardila2019-dm}, which provides speech samples from a larger number of unique speakers to ensure convergence to meaningful speaker embeddings without over-fitting. We evaluate on all LRL datasets.

For our BYOL approach, we train on both LRLs and HRLs in the BYOL objective and HRLs in the supervision objective simultaneously. We employ a 5-fold cross-validation strategy, where the left-out target language LRL fold serves as the test set, while the left-out HRL fold is used for validation.

We follow the same train/validation/test splits as provided in Common Voice and MSP-Podcast v1.11. For the other datasets, we adopt a leave-one-session-out cross-validation strategy, resulting in 5 CV splits 5-fold CV is applied over sessions, and averaged. We use the validation split for early stopping, model checkpoint selection, and hyperparameter tuning.



\subsection{Models Setup}
\textbf{CL.}
We build on Whisper’s encoder \cite{Radford2022-iu}, with three modifications: (1) Input length is reduced to 4 seconds by cropping the Positional Encoding layer, improving efficiency without performance loss. (2) Convolutional and positional encoding layers are frozen, following \cite{Osman2023-eu}, for more stable training. (3) A classification head with two fully connected layers and Dropout is added, using Whisper embeddings (average pooled over time) to generate a fixed-length utterance representation.

\textbf{BYOL.}
We use the Whisper’s encoder described in Section \ref{subsec:contrastive}, initialized from the HuggingFace checkpoint. Unlike the contrastive approach, the input length is capped at 3 seconds with cropped positional encoding. As in CL, convolutional and positional layers are frozen, and a two-layer classification head with Dropout is added to the time-averaged encoder outputs.

\textbf{Baseline. }
We use the same Whisper-based encoder as for CL, except that we initialize the model with different weights to explicitly isolate the impact of the speaker-contrastive adaptation stage.
The baseline is trained exclusively on English emotion-annotated datasets and evaluated on out-of-domain LRL datasets in a zero-shot manner.

\subsection{Training Details}

All models are implemented using PyTorch v2.5.1 and trained using a batch size of 64 with early stopping for computational efficiency, AdamW \cite{Loshchilov2017-ra} 
optimizer with a learning rate of 1e-5. For all experiments, a training epoch lasts for 100 batches. We use Whisper.small.en via transformers 4.39.3. We repeat all experiments five times and report the mean and standard deviation of our metrics. During BYOL training, we linearly scale down the BYOL loss factor $\lambda$ from $0.8$ to $0.2$. All trainings are executed on 2 Nvidia A5500 and 2 Nvidia L40 GPUs. 




\begin{table}[t]
    \centering
    \resizebox{.45\textwidth}{!}{
        \begin{tabular}{lcccc}
            \toprule
            \textbf{Dataset} & \textbf{Metric$_{(std)}$} & \textbf{Baseline} & \textbf{Contrastive} & \textbf{BYOL}  \\
            \midrule
            {\textbf{URDU}}  
                    & Accuracy  & 0.597$_{(0.025)}$ & 0.648$_{(0.031)}$ & \textbf{0.658$_{(0.052)}$} \\
            Urdu    & Macro F1  & 0.547$_{(0.036)}$ & 0.629$_{(0.041)}$ & \textbf{0.653$_{(0.060)}$} \\
                    & UAR       & 0.597$_{(0.025)}$ & 0.637$_{(0.032)}$ & \textbf{0.659$_{(0.053)}$} \\
            \midrule
            {\textbf{EMODB}}  
                    & Accuracy  & 0.776$_{(0.040)}$ & \textbf{0.906}$_{(0.017)}$ & 0.794$_{(0.033)}$  \\
            German  & Macro F1  & 0.750$_{(0.059)}$ & \textbf{0.901}$_{(0.018)}$ & 0.732$_{(0.030)}$  \\
                    & UAR       & 0.758$_{(0.050)}$ & \textbf{0.896}$_{(0.019)}$ & 0.751$_{(0.033)}$  \\
            \midrule
            {\textbf{SUBESCO}}  
                    & Accuracy  & 0.616$_{(0.015)}$ & \textbf{0.721}$_{(0.008)}$ & 0.643$_{(0.012)}$ \\
            Bangla  & Macro F1  & 0.572$_{(0.016)}$ & \textbf{0.711}$_{(0.011)}$ & 0.641$_{(0.009)}$ \\
                    & UAR       & 0.616$_{(0.015)}$ & \textbf{0.721}$_{(0.008)}$ & 0.643$_{(0.014)}$ \\
            \midrule
            {\textbf{RAVDESS}}
                    & Accuracy  & 0.574$_{(0.020)}$ & 0.580$_{(0.018)}$ & \textbf{0.582$_{(0.032)}$} \\
            English & Macro F1  & 0.514$_{(0.025)}$ & 0.519$_{(0.023)}$ & \textbf{0.561$_{(0.029)}$} \\
                    & UAR       & \textbf{0.578$_{(0.034)}$} & 0.570$_{(0.026)}$ & 0.572$_{(0.030)}$ \\
            \bottomrule
        \end{tabular}
    }
    \caption{Performance$_\text{(std)}$ Comparison of Baseline, CL and BYOL models on different datasets along with their language}
    \label{tab:performance}
    \vspace{-0.3cm}
\end{table}

\section{Results and Discussion} 
For all three LRL datasets, models leveraging contrastive learning and BYOL as a pre-training stage outperformed the baseline. Table \ref{tab:performance} reports Accuracy, Macro F1, and UAR for the three LRL datasets and a held-out HRL dataset. Results are averaged over five runs with the standard deviation in the footnote and with the best-performing models in bold. 

The largest improvement in CL was observed on the German dataset, with nearly a 20\% increase across all metrics. Notably, the Speaker-Contrastive model achieved 90\% zero-shot accuracy on EmoDB without direct training, marking the best-known performance on this dataset in an out-of-domain setting. This may be due to the dataset’s more distinguishable emotional content. However, we also hypothesize that German’s linguistic proximity to English may contribute, as our models undergo final fine-tuning on English speech-annotated data, potentially biasing them toward languages closer to English.

In contrast, the BYOL approach demonstrated its strength in other scenarios. On the URDU dataset, BYOL achieved the highest accuracy (65.8\%) and Macro F1 (65.3\%) scores among the three methods, outperforming both the baseline and the contrastive model. Similarly, for SUBESCO, BYOL also surpassed the baseline across all metrics, though the contrastive approach still led overall improvements. These results suggest that BYOL effectively learns robust latent representations in low-resource settings, providing a viable alternative when data is scarce. Yet, it is important to note that BYOL exhibit larger standard deviations, particularly on URDU and EmoDB, probably due to the small size of each cross-validation fold ($ < 100$ samples). 

The smallest gains in both approaches were observed on the English RAVDESS dataset, held out from training alongside URDU, EmoDB, and SUBESCO. Performance on RAVDESS remained nearly identical to the baseline, showing no measurable improvement but also no deterioration. This suggests that despite our approach’s modifications, the models retain their discriminatory ability in HRLs.

\begin{figure}
    \centering
    \includegraphics[width=1.0\linewidth]{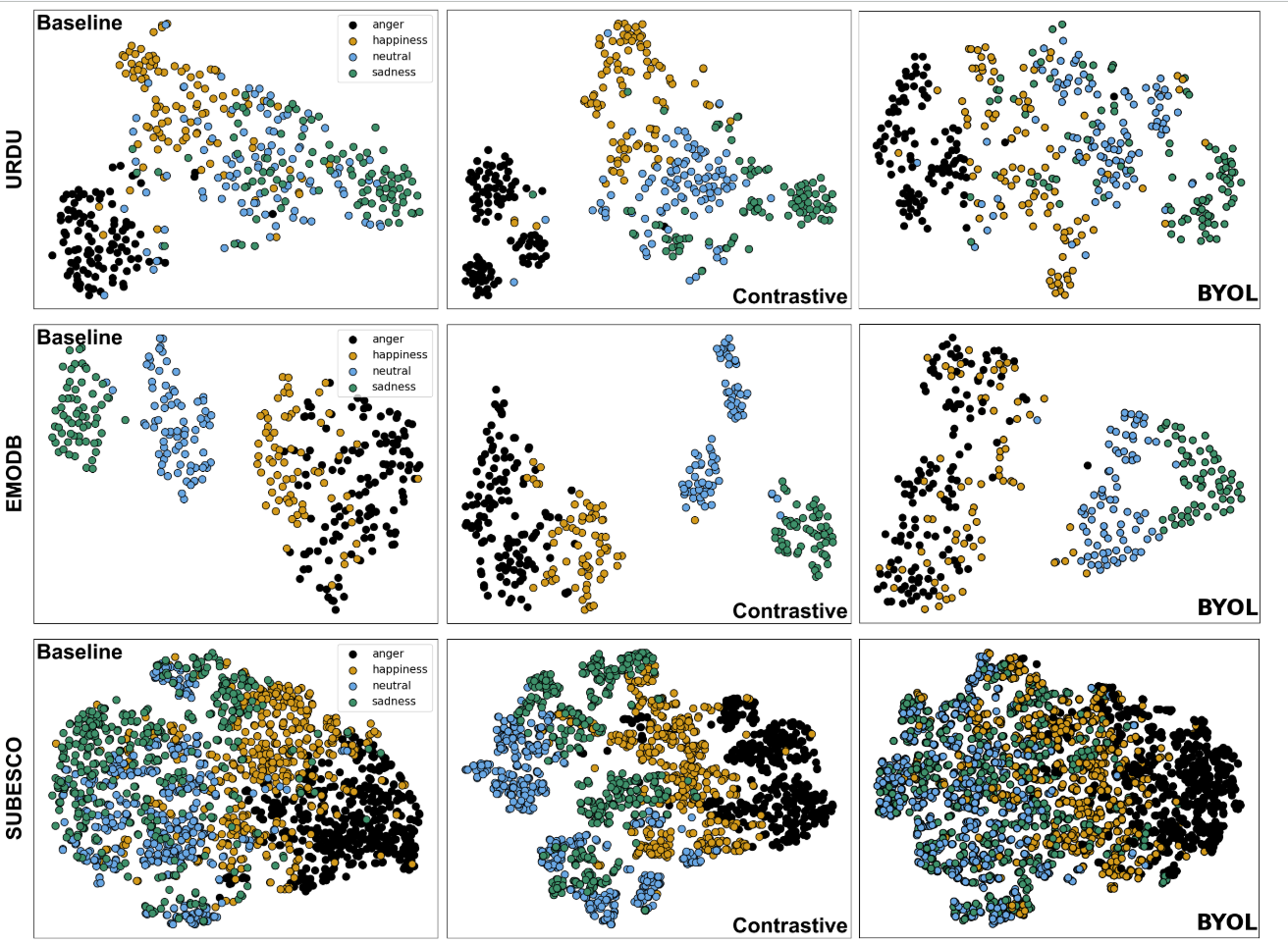}
    \caption{T-SNE plots of learned embeddings on LRLs. Left: Baseline model after training. Middle: Speaker-contrastive model after training. Right: BYOL model after training.}
    \label{fig:tsne}
    \vspace{-0.2cm}
\end{figure}

\subsection{Gender Bias in Urdu SER}
When we analyzed common failures of our models on the URDU dataset, we observed a strong gender bias in the CL model. The baseline model correctly predicts 74\% of female and 54\% of male speakers. While it achieved a 10\% improvement, the CL model regressed on female speech (41\%) while improving for males (73\%). Nevertheless, the BYOL approach has 55\% correctness on female speakers and 66\% correctness on male speakers. We hypothesize that BYOL mitigates gender bias, as it does not use speaker driven positive/negative pair construction during training. This imbalance, obscured by the dataset’s 86\% male dominance, highlights a fairness issue for future work and improvement in CL approach.

No similar bias was observed in other datasets, where the speaker-contrastive model improved accuracy equally across genders. We suspect that this issue stems from random speaker sampling during contrastive adaptation and the male-skewed Urdu Common Voice dataset. We believe a robust and fair SER model should minimize gender imbalances, and future work will focus on mitigating this issue.

\subsection{Interpreting Models Embeddings}

Understanding model behavior, as seen in the hidden gender bias example, is challenging. One common interpretability method is T-SNE visualizations, which helps analyze learned embedding spaces. Figure \ref{fig:tsne} compares embeddings from the baseline, CL models, and BYOL models across the three LRL SER datasets. The CL and BYOL models exhibit more distinct emotion class boundaries, compared to the baseline model, supporting our empirical findings of improved performance. Comparing the T-SNE embeddings of CL and BYOL, CL has more dense clusters and creates more space between emotion classes. Notably, stronger clustering appears between “anger”/“happiness” and “neutral”/“sad”. These pairs share similar arousal and dominance characteristics but differ in valence, which often depends on lexical content rather than speech patterns. In other words, distinguishing neutral from sad relies more on what is said than on how it is said. 

This lexical dependence poses challenges in multilingual transfer learning, where valence cues may not transfer across languages. A word or sound conveying negative sentiment in one language may carry a different emotional meaning in another. For example, the sound “ach” denotes negative valence in English but is strongly positive in German.
Our speaker-contrastive adaptation stage mitigates this issue by encouraging the encoder to prioritize speaker-distinguishing speech features rather than potentially misleading lexical cues. As shown in Table \ref{tab:confusion}, performance gains mainly result from fewer “anger”/“happiness” and “neutral”/“sad” misclassifications.

\subsection{Impact of Source Data in Self-Supervised Training}
We examined the impact of different data sources on self-supervised training in BYOL and CL. Our findings show that CL benefits from a larger, diverse LRL dataset, even if unlabeled, while BYOL performs better when trained on the target LRL dataset, especially when its distribution aligns with non-labeled LRL samples used in training. Specifically, we compared BYOL training with (i) the Common Voice subset + HRL and (ii) the target LRL dataset + HRL. While CL performance improved with the Common Voice subset, BYOL performed better when trained directly on the target LRL dataset. Conversely, using the target LRL dataset instead of Common Voice in CL led to a performance drop.
We hypothesize that this difference arises from how each method utilizes data. Since the Common Voice dataset is 10 times larger than the target LRL dataset, CL benefits from its greater speaker diversity, improving speaker-contrastive adaptation. In contrast, BYOL does not explicitly leverage speaker information, making the additional variety in the non-target LRL dataset less beneficial.

\begin{table}[t]
\centering
\resizebox{0.75\columnwidth}{!}{%
\begin{tabular}{@{}l|cccc@{}}
\toprule
\textbf{Truth\textbackslash{}Pred} & anger & happiness & neutral & sadness \\ \midrule
anger                          & +20   & -16       & -4      & 0       \\
happiness                      & 0     & -1        & -1      & +1      \\
neutral                        & 0     & -1        & -1      & +2      \\
sadness                        & 0     & 0         & -26     & +26     \\
\bottomrule
\end{tabular}
}
\caption{Confusion matrix for CL on EMODB, showing the mean performance difference between contrastive and baseline models. Higher diagonal values indicate better performance.}
\label{tab:confusion}
\vspace{-0.5cm}
\end{table}







\section{Conclusion and Limitations}
In this work, we show that Contrastive Learning and  and Bootstrap Your Own Latent enhance Speech Emotion Recognition performance in low-resource settings, with clearer emotion class separation in learned embeddings. Our contributions include \emph{i)} effective improvement in performance (F1) in low-resource settings: 10.6\% in Urdu, 15.2\% in German, 13.9\% in Bangla; \emph{ii)} providing deeper insights of model behavior through model analysis and interpretation, also identifying challenges in LRL emotion recognition. Despite these advances, our limitations include a focus on a limited number of languages, and potential biases introduced by pre-training data. 

Future work will explore linguistic distance in cross-lingual transfer, data augmentation to address gender imbalance, and modifications to CL and BYOL to reduce sensitivity to speaker-dependent features, improving embedding robustness. By addressing these challenges, we aim to advance fair and effective emotion recognition in diverse linguistic settings.

\section{Acknowledgements} 
This research is supported in part by the Defense
Advanced Research Projects Agency (DARPA), via
the CCU Program contract HR001122C0034, and the National Science Foundation via ARNI Columbia 2025 Research Project. The views, opinions and/or findings expressed are those
of the authors solely.


\bibliographystyle{IEEEtran}
\bibliography{mybib}

\end{document}